\documentstyle[12pt]{article}
\newcommand{\tr}[1]{\,{\rm tr}\,#1\,}
\newtheorem{lemma}{Proposition}
\newtheorem{theor}{Theorem}

\begin{document}
\title{
Geometric construction of the classical $R$-matrices for the elliptic and
trigonometric Calogero-Moser systems.}
\author{G.E.Arutyunov
\thanks{Steklov Mathematical Institute,
Vavilov 42, GSP-1, 117966, Moscow, Russia; arut@class.mian.su}\\
and\\
P.B.Medvedev \thanks
{Institute of Theoretical and Experimental Physics,
117259 Moscow, Russia}
}
\date {}
\maketitle
\begin{abstract}
By applying the Hamiltonian reduction scheme we recover the R-matrix
of the trigonometric and elliptic Calogero-Moser system.
\end{abstract}
\newpage
\section{Introduction}

Recently the theory of integrable many-body systems has a number of
interesting developments. In particular, classical dynamical R-matrices
were found for rational, trigonometric \cite{AT}, elliptic \cite{Skl}
Calogero systems and also for trigonometric Ruijsenaars-Schneider
model \cite{A}.

About twenty years ago it was found
\cite{OP}-\cite{OP2} that finite-dimensional dynamical systems of Toda- and
Calogero-types can be obtained by the Hamiltonian reduction of geodesic
motions on the cotangent bundles of semi-simple Lie groups.
The $L$-operator of a system
arises as a point on the reduced phase space ${\cal P}$,
whereas the Lax representation $\frac{dL}{dt}=[M,L]$ -  as the equation of
motion on ${\cal P}$.
Later on  it was realized \cite{FO} that this construction provides an
effective tool to deduce the classical
$R$-matrix and to prove thereby the
integrability of a system.  In the recent paper
by Avan, Babelon and Talon \cite{ABT} a nice
computation scheme along the lines described above was worked out in
detail and applied to recover the known \cite{AT} dynamical $R$-matrices
of the rational and trigonometric Calogero-Moser models.

The dynamical $R$-matrix for the elliptic Calogero model was found in
\cite{Skl} by the direct computation. In that paper a task to find a
geometric interpretation of this $R$-matrix was adduced (see also
\cite{ABT}).  The essential ingredient to solve this task, the Hamiltonian
reduction procedure, was elaborated in the recent paper by Gorsky and
Nekrasov \cite{GNH}. They proved that the phase space of the elliptic
Calogero model is the cotangent bundle to the moduli space of holomorphic
connections on a torus with a marked point.  We employ this reduction and
following the computation scheme of \cite{ABT} deduce the Sklyanin
$R$-matrix. It appears to be a solution of the first order differential
equation:  $$ X=[R(X),D] - k\bar{\partial} R(X)+Q, $$ where $D$ and $Q$
are constant diagonal matrices, and $X$ is an $sl(n,{\bf C})$ -valued
function on a torus.

The paper is organized as follows. In section 2
we show how the classical $R$-matrix is related to the factorization
problem for $su(n)$ connection on a circle (holomorphic $sl(n,{\bf C})$
connection on a torus). Sections 3 and 4 are devoted to the explicit
solution of the corresponding factorization problems.  As the result we
recover the trigonometric and elliptic $R$-matrices.

\section{$R$-matrix from the Hamiltonian reduction}
We start with the brief review of some basic facts about the Hamiltonian
reduction of cotangent bundles over affine algebras.

Denote by $G$ a finite-dimensional Lie group with a Lie algebra $\cal G$.
Let ${\cal L}{\cal G}=\{\phi:~S^1\rightarrow {\cal G}\}$ be a current
algebra and $\widehat{{\cal L}{\cal G}}$ be its central extension.
The commutation relations in $\widehat{{\cal L}{\cal G}}$ are
$$
[(\phi,c),(\phi',c')]=([\phi,\phi'], \int_{S^1}d\varphi \mbox{tr}
(\phi\partial \phi')),~~~c\in {\bf R}.
$$
The dual $\widehat{{\cal L}{\cal G}}^*$ to
$\widehat{{\cal L}{\cal G}}$ consists of
pairs $(A,k)$, where $A$: $S^1\rightarrow {\cal G}$ and $k\in{\bf R}$.
The nondegenerate pairing is
\begin{equation}
<(A,k),(\phi,c)>=\int_{S^1}\mbox{tr}(A\phi)+kc,
\label{sc}
\end{equation}
The sum $\widehat{{\cal L}{\cal G}}\oplus \widehat{{\cal L}{\cal G}}^*$
can be identified with
the cotangent bundle $T^*\widehat{{\cal L}{\cal G}}$ over
$\widehat{{\cal L}{\cal G}}$
supplied with the standard Poisson (symplectic) structure:
\begin{equation}
\{f,h\}=\int_{S^1}d\varphi \mbox{tr}\left(
\frac{\delta f}{\delta \phi}\frac{\delta h}{\delta A}-
\frac{\delta f}{\delta A}\frac{\delta h}{\delta \phi}
\right)+
\frac{\delta f}{\delta c}\frac{\delta h}{\delta k}-
\frac{\delta f}{\delta k}\frac{\delta h}{\delta c}.
\label{pois}
\end{equation}
Sometimes we will denote by $p$ the whole set $(\phi,c;A,k)$ being the
point of the phase space ${\cal P}$.

A current group ${\cal L}G$
acts on $\widehat{{\cal L}{\cal G}}$ and on  $\widehat{{\cal L}{\cal
G}}^*$ by the adjoint and coadjoint actions respectively:
\begin{equation}
(\phi,c)\rightarrow (g\phi g^{-1},~c+ \int_{S^1}d\varphi
(\phi g^{-1}\partial g)), \label{sc1}
\end{equation}
\begin{equation}
(A,k)\rightarrow
(gAg^{-1}-k\partial gg^{-1},~k).
\label{sc2}
\end{equation}
This action is Hamiltonian and it gives rise to a moment map
$\mu:~T^*\widehat{{\cal L}{\cal G}}\rightarrow {\cal L}{\cal G}$
\begin{equation}
\mu(p)=k\partial \phi+[A,\phi].
\label{momen}
\end{equation}
Let $J\in \widehat{{\cal L}{\cal G}}^*$. Then
a quotient $\mu^{-1}(J)/G_J$
by the action of the isotropy group $G_J\in G$  of $J$
admits under some natural assumptions a symplectic structure \cite{Arn}.

Let $A$ be a smooth function on $S^1$ with values in some real
semi-simple Lie algebra. A differential equation
\begin{equation}
A=-k\partial ff^{-1}
\label{mon}
\end{equation}
has a {\it unique} solution if one fixes $f(0)=1$. Since $A$ is periodic,
then $f(\varphi+2\pi)=f(\phi)M(A)$, where $M(A)$ is a constant matrix
called the monodromy of $A$. Thus we have a mapping $M$: $A\rightarrow M(A)$.
Performing the gauge transformation
\begin{equation}
A^{g}=g A g^{-1}-k\partial gg^{-1}
\label{bon}
\end{equation}
with an element $g$ of the current group $(g(0)=g(2\pi))$,
one finds that $M(A^{g})=g(0)M(A)g(0)^{-1}$. The inverse assertion is
also true, i.e. if we have two fields $A$ and $\tilde{A}$ with
monodromies related as $M(\tilde{A})=gM(A)g^{-1}$, then the function
$$
g(\varphi)=\tilde{f}(\varphi)gf^{-1}(\varphi)
$$
is a gauge transform from $A$ to $\tilde{A}$. Thus,
the spectral invariants of $M(A)$ completely parametrize the orbits of
(\ref{bon}).

Let us diagonalize $M(A)$: $M(A)\rightarrow {\cal D}(A)$ by some gauge
transformation of $A$ and define $D$ as $\exp{D}={\cal D}(A)$.
Now consider the function $f(\varphi)=e^{\frac{\varphi}{2\pi}D}$.
On the one hand, it is the solution of (\ref{mon}) for the constant
diagonal field $\tilde{A}=-\frac{k}{2\pi}D$ and on the other hand, it
defines the monodromy conjugated to $M(A)$. Thus every field $A$ is gauge
equivalent to a constant matrix with values in the Cartan subalgebra.
Moreover, if we fix the order of eigenvalues of $D$ (the fundamental Weyl
chamber), then $D$ is uniquely defined. Consequently we have a
decomposition
\begin{equation}
A=g(A,k)D(A,k)g(A,k)^{-1}-k\partial g(A,k) g(A,k)^{-1},
\label{dec}
\end{equation}
where $D(A,k)$ is the constant diagonal matrix with a fixed order
of eigenvalues.

Recall that in the approach of \cite{GN,GN1} the moment map
is fixed by the condition
\begin{equation}
k\partial \phi+[A,\phi]=
\nu \sum_{\alpha\in \Delta_{+}} (e_{\alpha}+e_{-\alpha})\delta(\varphi),
\label{gnc}
\end{equation}
where $\nu$ is a coupling constant.
The isotropy group of this moment is also large enough to reduce $A$ to
a constant diagonal matrix $D=iX$. Then solving (\ref{gnc}) with respect
to $\phi$, one gets \cite{GN} the $L$-operator of the trigonometric
Calogero model \cite{OP}:
\begin{equation}
L(\varphi)=iP-\frac{i\nu}{2k}\sum_{\alpha\in\Delta_{+}}
\left(
\frac{e^{-\frac{i\alpha(X)}{k}\varphi}}
{e^{-\frac{2\pi i}{k}\alpha(X)}-1}e_{\alpha}+
\frac{e^{\frac{i\alpha(X)}{k}\varphi}}
{e^{\frac{2\pi i}{k}\alpha(X)}-1}e_{-\alpha}
\right).
\label{rh}
\end{equation}
A pair $(L,D)$ is a point of the reduced phase space ${\cal P}_{\mu}$.
For the $su(n)$ case (\ref{rh}) reduces to
\begin{equation}
L(\varphi)=i\sum_{i}p_i E_{ii}+\frac{\nu}{2k}
\sum_{i\neq j}\frac{e^{\frac{i}{k}\alpha_{ij}(X)(\pi-\varphi)}}
{\sin{\frac{\pi}{k}\alpha_{ij}(X)}}E_{ij},
\label{op}
\end{equation}
where $E_{ij}$ is the basis of matrix unities. The description of the
reduction completes by mentioning that the entries of the constant
diagonal matrices $P$ and $X$ form a set of the canonically conjugated
variables being the coordinates on ${\cal P}_{\mu}$.

Now we proceed further with our analysis of (\ref{dec}). In contract to
$D(A,k)$, the factor $g(A,k)$ is not uniquely defined. In the case of a
simple real Lie group $G$ this ambiguity is described in the following
\begin{lemma} Factor $g(A,k)$ in (\ref{dec}) is defined up to the
right multiplication by elements of the maximal torus $T$ of $G$.
\end{lemma}
The proof
is as follows. Suppose that we have two elements $g$ and $g'$ of ${\cal L}G$
such that $A=g\circ D=g'\circ D$, where $\circ$ is a shorthand for
(\ref{dec}).  Then $g'^{-1}g\circ D=D$, i.e.  $\gamma=g'^{-1}g$ belongs to
the stabilizer of $D$, so that $\gamma(0)$ lies in the centralizer of
$M(D)$.  Since $M(D)$ is a generic diagonal matrix, we have $\gamma(0)\in
T$. It is readily seen that $\gamma(\varphi)$ is given by
$\gamma(\varphi)=f(\varphi)\gamma(0)f(\varphi)^{-1}$, where $f(\varphi)$
is the solution of (\ref{mon}) for $A=D$. Recalling that
$f(\varphi)=e^{-\frac{\varphi}{k}D}$, we get
$\gamma(\varphi)=\gamma(0)\in T$, i.e. $g'^{-1}g\in T$.

Let us associate with any element $X\in {\cal L}{\cal G}$ a function
\begin{equation}
F_X((\phi,c),(A,k))=<\phi,g(A,k)Xg(A,k)^{-1}>,
\end{equation}
where $g(A,k)$ is defined by (\ref{dec}).
The isotropy group $G_J$ acting
on the surface $\mu^{-1}(J)$ coincides with a group of smooth
mappings \mbox{$G_J=\{g:~S^1\rightarrow G,~~g(0)\in H\}$},  where $H$
is the isotropy group of
$J=\sum_{\alpha\in \Delta_{+}} (e_{\alpha}+e_{-\alpha})$.
Since $H\cap T=0$, there is no ambiguity
in the choice of $g(A,k)$ for $A$ lying on the surface $\mu^{-1}(J)$,
so that $F_X$ is well defined. Moreover, $F_X$ is invariant with respect
to (\ref{bon}) with $g\in G_J$, i.e. it is a function on the reduced phase
space.

Denote by $\xi_X$ the hamiltonian vector field corresponding to
$X\in {\cal L}{\cal G}$. In the sequel, we need to
know the action of $\xi_X$ on $F_Z$. The calculation is
straightforward
$$
\xi_X F_Z =\frac{d}{dt} F_Z(p(t))_{|_{t=0}}=\frac{d}{dt}
<e^{tX}\phi e^{-tX}, g(e^{tX}\circ (A,k))Zg(e^{tX}\circ
(A,k))^{-1}>_{|_{t=0}}
$$
$$ =<\phi,
g(A,k)\left[g(A,k)^{-1}\nabla_{(A,k)}g(X)-g(A,k)^{-1}Xg(A,k),Z\right]
g(A,k)^{-1}>,
$$
where the derivative along an orbit of gauge transformations
\begin{equation}
\nabla_{(A,k)}g(X)=\frac{d}{dt}g(e^{tX}\circ (A,k))_{|_{t=0}}
\label{der}
\end{equation}
was introduced.
On ${\cal P}_{\mu}$ the previous formula takes the form
\begin{equation}
\xi_X F_Z = <L,\left[\nabla_{(D,k)}g(X)-X,~Z \right]>.
\label{kl}
\end{equation}
As it was explained in \cite{ABT}, the Poisson bracket on the reduced
phase space can be presented in the following convenient form
\begin{equation}
\{f,h\}_r=\{f,h\}-<J,[V_f,V_h]>,
\label{brac}
\end{equation}
where $f,h$ are functions on ${\cal P}$ whose restrictions on
$\mu^{-1}(J)$ are invariant with respect to the action of the isotropy
group and $V_f$ is the solution of $<J,[X,V_{f}]>=\xi_X f$ (see \cite{ABT}
for details).  Since $F_X, F_Y$ are invariant on $\mu^{-1}(J)$,
we get
\begin{equation}
\{F_X,F_Y\}_r=\{F_X,F_Y\}-<J,[V_{F_X},V_{F_Y}]>,
\label{dsa}
\end{equation}
The second term in (\ref{dsa}) follows from (\ref{kl}):
$$
<J,[V_{F_X},V_{F_Y}]>=
<L,\left[\nabla_{(D,k)}g(V_{F_X})-V_{F_X},~Y \right]>.
$$
To obtain the bracket $\{F_X,F_Y\}$ we first calculate
$$
\frac{\delta F_X}{\delta A_{ij}(\varphi')}=
<\phi, g(A)\left[g(A)^{-1}\frac{\delta g(A)}{\delta A_{ij}(\varphi')},
X \right]g(A)^{-1}>,
$$
$$
\frac{\delta F_X}{\delta \phi_{ij}(\varphi')}=
(g(A)Xg(A)^{-1})_{ij}(\varphi')
$$
and then substitute the result in (\ref{pois}):
\begin{equation}
\{F_X,F_Y\}=
<\phi, g(A)\left
[g(A)^{-1}\int_{S^1}(g(A)Xg(A)^{-1})_{ij}(\varphi')
\frac{\delta g(A)}{\delta A_{ij}(\varphi')},Y \right]g(A)^{-1}>
\label{brb}
\end{equation}
$$
-<\phi, g(A)\left
[g(A)^{-1}\int_{S^1}(g(A)Yg(A)^{-1})_{ij}(\varphi')
\frac{\delta g(A)}{\delta A_{ij}(\varphi')},X \right]g(A)^{-1}>.
$$
Thus we have
\begin{lemma}
There exists a linear operator $R$: ${\cal G}\rightarrow {\cal G}$
given by
\begin{equation}
R(X)(\varphi)=\sum_{ij}\int_{S^1}d\varphi'X_{ij}(\varphi')
\frac{\delta g(D,k)}{\delta A_{ij}(\varphi')}(\varphi)
-\frac{1}{2}(\nabla_{(D,k)}g(V_{F_X})-V_{F_X})
\label{vbas}
\end{equation}
such that
the Poisson bracket on the reduced phase space is of the form
\begin{equation}
\{F_X,F_Y\}_{r}=
<L,[R(X),Y]+[X,R(Y)]>
\label{rt}
\end{equation}
\end{lemma}
{\it Remark.} Proposition 2 is a generalization of Theorem 4.1 of
\cite{ABT} to the reduction procedure described above.

$R$-matrix (\ref{vbas}) depends on the extension of $F_X$ in the vicinity
of $\mu^{-1}(J)$. We extend $F_X$ in a specific way.

For the sake of
being definite, assume that ${\cal G}=su(n)$. Then
${\cal G}$ has the decomposition in the direct sum
${\cal G}={\cal H}\oplus {\cal B}\oplus {\cal C}$,
where ${\cal H}$ is a Lie algebra of the isotropy group $H$, $\cal B$ is
the Lie algebra of $T$ and $\cal C$ is defined as an orthogonal
to ${\cal H}\oplus {\cal B}$ with respect to the Killing metric.
Consider the factor $g(A,k)=\exp{X}$ in (\ref{dec}), where $X\equiv
X(\varphi)$ is an element of the current algebra. Using the ambiguity in
the definition of $g(A,k)$ described by Proposition 1, we arrive at
\begin{lemma}
For any pair $(A,k)$ with $k\neq 0$ decomposition (\ref{dec})
defines a unique element $g(A,k)=e^{X(\varphi)}$ of the current group
such that $X(\varphi)$ is an element of the current algebra
obeying the boundary condition  $X(0)\in {\cal H}\oplus {\cal C}$. In
addition, if $(A,k)\in \mu^{-1}(J)$, then  $X(0)\in {\cal H}$.
\end{lemma}

Now return to (\ref{vbas}) and choose the extension of $g(A,k)$ as described
in Proposition 3.
Then $\nabla_{(D,k)}g(X)=P_{{\cal H}\oplus {\cal C}}$, where
$P_{{\cal H}\oplus {\cal C}}$ is a projector on
${\cal H}\oplus {\cal C}$ parallel to $\cal B$. Now
if $X(\varphi)$ is such that $X(0)\in {\cal H}\oplus {\cal C}$, then
$$
0=<L,\nabla_{(D,k)}g(X)-X>=<J,[X,V_{F_Y}]>.
$$
Since ${\cal H}\oplus {\cal C}$ is an isotropic space of the form
$\Xi(X,Y)=<J,[X,Y]>$ \cite{ABT}, we conclude that
$V_{F_Y}\in {\cal H}\oplus {\cal C}$ for any $Y$.
Hence under the choice of $g(A,k)$ given by Proposition 3 $R$-matrix
(\ref{vbas}) reduces to
\begin{equation}
R(X)(\varphi)=\sum_{ij}\int_{S^1}d\varphi'X_{ij}(\varphi')
\frac{\delta g(D,k)}{\delta A_{ij}(\varphi')}(\varphi).
\label{bv}
\end{equation}
The last formula has a transparent geometric meaning. Defining the
time evolution of the field $A(t)$ such that $A(0)=D$ and
$\frac{dA}{dt}_{|_{t=0}}=X$,
one has  $R(X)(\varphi)=\frac{d}{dt}g(A(t))(\varphi)_{|_{t=0}}$. Since
eq.(\ref{dec}) is valid for any $t$
$$
A(t)=g(A,k)(t)D(A,k)(t)g(A,k)(t)^{-1}-k\partial g(A,k)(t) g(A,k)(t)^{-1},
$$
we differentiate it with respect to $t$ and put $t=0$. The result is
\begin{equation}
X=[R(X), D]-k \partial R(X)+ Q,
\label{cet}
\end{equation}
where $Q=\frac{d}{dt}D_{|_{t=0}}$. Eq.(\ref{cet}) is a differential
equation of the first order. For any smooth function $X(\varphi)$ on
a circle it has a unique solution $R(X)$ obeying the boundary condition
$X(0)\in {\cal H}\oplus {\cal C}$. From (\ref{cet}) we can also
read off that the $R$-matrix is dynamical since it depends on $D$ that
accumulates the coordinates on the reduced phase space.  In the sequel,
we refer to (\ref{cet}) as to the factorization problem for $su(n)$
($sl(n,{\bf C})$) connection.  Hence, by construction the $R$-matrix
of the trigonometric Calogero model is
defined as a unique solution of the factorization problem for $su(n)$
connection obeying some specific boundary condition.
\section{Trigonometric case}
In this section we solve explicitly the factorization problem for
$su(n)$ connection and thereby recover the $R$-matrix of the
trigonometric Calogero model first found in \cite{AT}.

We start with equation
\begin{equation}
V=[\Lambda ,D]-k\Lambda '+t,~~~V,\Lambda \in {\cal L}su(n)
\label{wq}
\end{equation}
and $D,t$ are constant diagonal matrices.
Writing down the root decomposition of $su(n)$ elements
$$
V=\sum_{\alpha\in\Delta_+}v_{\alpha}e_{\alpha}-\bar{v}_{\alpha}e_{-\alpha}+
\sum_{i}v_ih_i,~~~
\Lambda
=\sum_{\alpha\in\Delta_+}x_{\alpha}e_{\alpha}-\bar{x}_{\alpha}e_{-\alpha}
+\sum_{i}x_ih_i,
$$
and introducing $v_{\alpha}=<V,e_{-\alpha}>$, $v_i=<V,h_i>$, ets.,
we get from (\ref{wq}) two equations on diagonal and nondiagonal parts of
$\Lambda $ respectively.
Imposing on $\Lambda $ the periodicity condition: $\Lambda (0)=
\Lambda (2\pi)$, we reconstruct $\Lambda $ up to
a constant diagonal matrix $h$ with pure
imaginary entries:
\begin{equation}
\Lambda (\varphi)=
\frac{\varphi}{2\pi k}
\int_{0}^{2\pi}v_i h_i-\frac{1}{k}\int_{0}^{\varphi}v_ih_i+h
\label{arbr}
\end{equation}
$$
\frac{i}{2k}
\sum_{\alpha\in\Delta_{+}}
\left(
\frac{e^{-\frac{i\pi\alpha(X)}{k}}}
{\sin\frac{\pi}{k}\alpha(X)}
\int_{0}^{2\pi}d\theta
e^{-i\frac{\alpha(X)}{k}(\varphi-\theta)}v_{\alpha}(\theta)e_{\alpha}+
\frac{e^{\frac{i\pi\alpha(X)}{k}}}
{\sin\frac{\pi}{k}\alpha(X)}
\int_{0}^{2\pi}d\theta
e^{i\frac{\alpha(X)}{k}(\varphi-\theta)}\bar{v}_{\alpha}(\theta)e_{-\alpha}
\right)
$$
$$
-\frac{1}{k}\sum_{\alpha\in\Delta_{+}} \left(
\int_{0}^{\varphi}d\theta
e^{-i\frac{\alpha(X)}{k}(\varphi-\theta)}v_{\alpha}(\theta)e_{\alpha}-
\int_{0}^{\varphi}d\theta
e^{i\frac{\alpha(X)}{k}(\varphi-\theta)}\bar{v}_{\alpha}(\theta)e_{-\alpha}
\right),
$$
where we put $D=iX$.

Now we fix $h$ by
requiring $\Lambda (0)$  to be an element of ${\cal H}\oplus {\cal C}$.
To this end we choose the explicit realization of the root basis
by the matrix unities and  evaluate
$\Lambda (\varphi)$ in (\ref{arbr}) at zero point:
\begin{equation}
\Lambda (0)=\frac{i}{2k}\sum_{\i\neq j}
\frac{e^{\frac{i\pi \alpha_{ij}(X)}{k}}}
{\sin{\frac{\pi}{k}\alpha_{ij}(X)}}
\int_{0}^{2\pi}e^{i\frac{\alpha_{ij}}{k}\theta}v_{ij}E_{ij}+h.
\label{nu}
\end{equation}
In (\ref{nu}) the convenient notation $v_{ji}=-\bar{v}_{ij}$ was used.
Recall that any element $\Lambda (0)\in {\cal H}\oplus {\cal C}$ should
satisfy the relation (see \cite{ABT} for the proof)
\begin{equation}
\mbox{Im}\sum_{j}(\Lambda ((0))_{ij}
=\frac{1}{n}\sum_{i\neq j}\mbox{Im}(\Lambda (0))_{ij}
.\label{rd}
\end{equation}
Substituting
(\ref{nu}) in (\ref{rd}), one finds $h$ that makes (\ref{rd}) true:
\begin{equation}
h=\frac{i}{4k}\sum_{\i\neq j}
\frac{e^{\frac{-i\pi \alpha_{ij}(X)}{k}}}
{\sin{\frac{\pi}{k}\alpha_{ij}(X)}}
\int_{0}^{2\pi}e^{i\frac{\alpha_{ij}}{k}\theta}v_{ij}
\left((\frac{1}{n}-E_{ii})+(\frac{1}{n}-E_{ii})\right).
\label{nu1}
\end{equation}
Combining (\ref{arbr}) and (\ref{nu1}), we finally get
\begin{lemma}
The solution of the factorization problem (\ref{wq}) obeying the constraint
$\Lambda (0)\in {\cal H}\oplus {\cal C}$ has the form:
\begin{equation}
\Lambda (\varphi)=\frac{\varphi}{2\pi k}\int_{0}^{2\pi}v_iE_{ii}-
\frac{1}{k}
\int_{0}^{\varphi}v_iE_{ii}+
\label{solu}
\end{equation}
$$
\frac{i}{2k}\sum_{i\neq j}
\frac{e^{-\frac{i\pi \alpha_{ij}(X)}{k}}}
{\sin{\frac{\pi\alpha_{ij}(X)}{k}}}
\int_{0}^{2\pi}
e^{-\frac{i\alpha_{ij}(X)}{k}(\varphi-\varphi')}v_{ij}~E_{ij}
-\frac{1}{k}\sum_{i\neq j}
\int_{0}^{\varphi}
e^{-\frac{i\alpha_{ij}(X)}{k}(\varphi-\varphi')}v_{ij}~E_{ij}+
$$
$$
\frac{i}{4k}
\sum_{i\neq j}\left(
\frac{e^{-\frac{i\pi \alpha_{ij}(X)}{k}}}
{\sin{\frac{\pi\alpha_{ij}(X)}{k}}}
\int_{0}^{2\pi}
e^{\frac{i\alpha_{ij}(X)}{k}\varphi'}v_{ij}
\right)
\left(
(\frac{1}{n}-E_{ii})+(\frac{1}{n}-E_{jj})
\right).
$$
\end{lemma}
Clearly, we can rewrite it as
$$
\Lambda (\varphi)=\frac{\varphi-\pi}{2\pi k}\int_{0}^{2\pi}v_iE_{ii}
+\frac{i}{2k}\sum_{i\neq j}
\frac{\cos{\frac{\pi\alpha_{ij}(X)}{k}}}
{\sin{\frac{\pi\alpha_{ij}(X)}{k}}}
\int_{0}^{2\pi}
e^{-\frac{i \alpha_{ij}(X)}{k}(\varphi-\varphi')}v_{ij}~E_{ij}
$$
$$
\frac{i}{4k}
\sum_{i\neq j}\left(
\frac{e^{-\frac{i\pi \alpha_{ij}(X)}{k}}}
{\sin{\frac{\pi\alpha_{ij}(X)}{k}}}
\int_{0}^{2\pi}
e^{\frac{i \alpha_{ij}(X)}{k}\varphi'}v_{ij}
\right)
\left(
(\frac{1}{n}-E_{ii})+(\frac{1}{n}-E_{jj})
\right)
$$
$$
+\frac{1}{2k}
\int_{0}^{2\pi}
\left(\sum_{i\neq j}
e^{-\frac{i
\alpha_{ij}(X)}{k}(\varphi-\varphi')}v_{ij}E_{ij}+\sum_{i}v_iE_{ii}\right)
\epsilon (\varphi-\varphi'),
$$
where the function
$$
\epsilon (\varphi-\varphi')=[1-2\theta(\varphi-\varphi')]=
\left\{
\begin{array}{l}
1,~~\mbox{if $\varphi\geq\varphi'$},\\
-1,~~\mbox{otherwise}
\end{array} \right.
$$
was used.

{}From this form of the solution we can read off that
$R$-matrix of the trigonometric Calogero system is a function
on $S^1\times S^1$ having the following explicit form
\begin{equation}
R(\varphi,\varphi')=\frac{\varphi-\pi}{2\pi k}\sum_{i}E_{ii}\otimes E_{ii}
+\frac{i}{2k}\sum_{i\neq j}
\frac{\cos{\frac{\pi\alpha_{ij}(X)}{k}}}
{\sin{\frac{\pi\alpha_{ij}(X)}{k}}}
e^{-i\frac{\alpha_{ij}(X)}{k}(\varphi-\varphi')}~E_{ij}\otimes E_{ij}
\label{dum}
\end{equation}
$$
-\frac{1}{2k}\sum_{i\neq j}
\left(E_{ii}-\frac{1}{n}\right)\otimes
\left(
\frac{e^{-i\frac{\alpha_{ij}(X)}{k}}}
{1-e^{-\frac{2\pi i\alpha_{ij}(X)}{k}}}E_{ij}
-
\frac{e^{i\frac{\alpha_{ij}(X)}{k}}}
{e^{\frac{2\pi i\alpha_{ij}(X)}{k}}}E_{ji}\right)
+\frac{1}{2k}r(\varphi,\varphi'),
$$
where we have introduced a matrix $r$:
\begin{equation}
r(\varphi,\varphi')=
\left(
\sum_{i\neq j}
e^{-\frac{i\alpha_{ij}(X)}{k}(\varphi-\varphi')}
E_{ij}\otimes E_{ji}+\sum_{i}E_{ii}\otimes E_{ii}\right)
\epsilon (\varphi-\varphi'),
\label{addi}
\end{equation}

It holds the following
\begin{lemma}
Matrix $r$ leads to the trivial Poisson bracket on the reduced phase space,
i.e. the following relation is satisfied
\begin{equation}
[r_{12}(\varphi,\varphi'), L(\varphi)\otimes I]-
[r_{21}(\varphi',\varphi), I\otimes L(\varphi')]=0.
\label{qaq}
\end{equation}
\end{lemma}
See Appendix A for the proof.
Consequently,
\begin{lemma}
$R$-matrix of the trigonometric Calogero model is given by (\ref{dum})
with $r(\varphi,\varphi')=0$.
\end{lemma}
{\it Remark.} On the reduced phase space the variables
$(P,X)$ are canonically conjugated. In standard tensor notation,
\begin{equation}
\{P_1,X_2\}=-\frac{1}{2\pi}\sum_{i} E_{ii}\otimes E_{ii}.
\label{can}
\end{equation}
$L$-operator (\ref{op}) as well as $R$-matrix (\ref{dum}) depend
on the phase $\varphi$. However, this dependence may be removed by
the similarity transformation
$L\rightarrow \tilde{L}=Q(\varphi)L(\varphi)Q(\varphi)^{-1}$, where
$Q(\varphi)=e^{-\frac{i}{k}X(\pi-\varphi)}$. The modified $L$-operator:
$$
\tilde{L}=i\sum_{i}p_i E_{ii}+\frac{\nu}{2k}
\sum_{i\neq j}
\frac{E_{ij}}{\sin{\frac{\pi}{k}\alpha_{ij}(X)}}
$$
with $(P,X)$ subjected to (\ref{can}) has also an $R$-matrix bracket
with
$$
R=\frac{i}{2k}\left(
\sum_{i\neq j}
\frac{\cos{\frac{\pi\alpha_{ij}(X)}{k}}}
{\sin{\frac{\pi\alpha_{ij}(X)}{k}}}~E_{ij}\otimes E_{ij}
+\frac{1}{2}\sum_{i\neq j}
\frac{1}{\sin{\frac{\pi\alpha_{ij}(X)}{k}}}
\left(E_{ii}-\frac{1}{n}\right)\otimes
(E_{ij}-E_{ji})\right).
$$
This $R$-matrix was first found in \cite{AT} and then recovered in
\cite{ABT} by the Hamiltonian reduction applied to the cotangent bundle
$T^*G$ over a finite-dimensional simple Lie group $G$.

\section{Elliptic case}
As it was first shown in \cite{GNH} the phase space of the elliptic
Calogero model coincides with the moduli space of
holomorphic connections on a torus $\Sigma_{\tau}$ with a marked point.
We employ this construction to deduce the classical $R$-matrix of the
elliptic Calogero model \cite{Skl}.

This time the phase space is characterized by the set
${\cal P}=(\phi,c;A,k)$, where $\phi,A$ are functions on $\Sigma_{\tau}$
with values in $sl(n,{\bf C})$, $c,k\in{\bf C}$. ${\cal P}$ can be
identified with the cotangent bundle over the centrally extended
current algebra $(\phi,c)$ of $sl(n,{\bf C})$-valued functions on
$\Sigma_{\tau}$. On ${\cal P}$ there is an
action of the current group ${\cal L}SL(n,{\bf C})$
\begin{eqnarray}
(\phi(z,\bar{z}),c) &\rightarrow &
(g(z,\bar{z})\phi(z,\bar{z})g^{-1}(z,\bar{z}),~~
c+\int_{\Sigma_{\tau}}d\bar{\eta} d\eta~\tr{\phi A}),
\label{ab}\\
(A(z,\bar{z}),k) &\rightarrow &
(g(z,\bar{z})A(z,\bar{z})g^{-1}(z,\bar{z})
-k\bar{\partial}g(z,\bar{z})g^{-1}(z,\bar{z}),k)
\nonumber \\
\end{eqnarray}
that preserves the standard symplectic structure (\ref{pois}).
The moment map of this action is fixed to be
\begin{equation}
k\bar{\partial}\phi+[A,\phi]=\nu J \delta(z,\bar{z})
\label{const}
\end{equation}
that defines the phase space and the $L$-operator of the elliptic
Calogero model \cite{GNH,G}. In (\ref{const}) $J$ denotes some
element on the coadjoint $sl(n,{\bf C})$ orbit.

Retracing the same steps as in the trigonometric case we can prove
without problems that the $R$-matrix corresponding to the $L$-operator
arising from (\ref{const}) is given by the similar formula
\begin{equation}
R(V)(z,\bar{z})=\sum_{ij}\int_{\Sigma_{\tau}}d\bar{\eta} d\eta~
V_{ij}(\eta,\bar{\eta})\frac{\delta g(A,k)}{\delta A_{ij}(z,\bar{z})}
(\eta,\bar{\eta})=X(z,\bar{z}),
\label{matr}
\end{equation}
where $X(z,\bar{z})$ is a solution of the factorization problem
\begin{equation}
V(z,\bar{z})=[X(z,\bar{z}),D]-k\bar{\partial}X(z,\bar{z})+t,
\label{fac}
\end{equation}
for $sl(n,{\bf C})$ connection $V$.
Here $D$ and $t$ are constant diagonal matrices, and
$V(z,\bar{z})$ and $X(z,\bar{z})$ are $sl(n,{\bf C})$-valued functions on
$\Sigma_{\tau}$. As usual we regard functions on a torus as twice periodic
functions on $\bf C$. From (\ref{fac}) it is easy to see that
$X(z,\bar{z})$ is defined only up to a constant diagonal matrix $h$.
As it was shown above $h$ must be fixed in a specific way to make (\ref{matr})
correct.

We start the study of (\ref{fac}) with solving the equation
\begin{equation}
\bar{\partial}x(z,\bar{z})=f(z,\bar{z}),
\label{sol}
\end{equation}
where $f(z,\bar{z})$ is a twice periodic function on $\bf C$ with periods
$\tau$ and $\tau'=1$. Suppose that we have a solution $x(z,\bar{z})$ of
(\ref{sol}), then $x(z+\tau,\bar{z}+\tau)$ is also a solution. Indeed,
substituting $z=w+\tau$ in (\ref{sol}), we get
$$
\frac{\partial}{\partial
\bar{w}}x(w+\tau,\bar{w}+\bar{\tau})=
f(w+\tau,\bar{w}+\bar{\tau})=f(w,\bar{w}).
$$
Consider the difference $\psi=x(w+\tau,\bar{w}+\bar{\tau})-x(w,\bar{w})$.
It satisfies the homogeneous equation $\bar{\partial}\psi=0$, i.e. $\psi$ is
an entire function.
In this way we found the monodromy property of a general solution of
(\ref{sol}):
\begin{equation}
x(w+\tau,\bar{w}+\bar{\tau})=x(w,\bar{w})+\psi_{\tau}(w).
\label{sol1}
\end{equation}

Suppose we have an equation on a torus
\begin{equation}
\bar{\partial}{\cal E}(z,\bar{z})=\delta(z,\bar{z}).
\label{sol2}
\end{equation}
In the vicinity of the origin eq.(\ref{sol2}) defines
a meromorphic function with a first order pole with
the residue $1/2\pi i$. Define a solution of (\ref{sol2}) as a
meromorphic function on $\bf C$ having simple poles
at the points ${\bf Z}\tau+{\bf Z}\tau'$ with the residues
$1/2\pi i$ and satisfying the quasiperiodicity conditions
\begin{equation}
{\cal E}(z+\tau,\bar{z}+\bar{\tau})={\cal E}(z,\bar{z})+C_{\tau},
\label{ee}
\end{equation}
$$
{\cal E}(z+\tau ',\bar{z}+\bar{\tau '})={\cal E}(z,\bar{z})+C_{\tau '},
$$
where $C_{\tau}, C_{\tau '}$ are complex numbers. As it will be seen only
these fundamental solutions are relevant to define a solution of
(\ref{sol}).

{\it Remark.} Note that ${\cal E}_{C_{\tau},C_{\tau'}}(z)$
can not be twice periodic since there is no elliptic functions of the
first order.  Here the subscript marks the monodromy property of the
solution.
\vskip 0.5cm

Suppose that we have two solutions ${\cal E}_{C_{\tau_1},C_{\tau'_1}}(z)$
and ${\cal E}_{C_{\tau_1},C_{\tau'_1}}(z)$. Their difference is an entire
but nonperiodic function $\psi$ (poles and residues of ${\cal E}$'s
coincide) with
\begin{equation}
\psi(z+\tau)=\psi(z)+C_{\tau_1}-C_{\tau_2}=\psi(z)+\delta,
\label{as}
\end{equation}
$$
\psi(z+\tau')=\psi(z)+C_{\tau_1'}-C_{\tau_2'}=\psi(z)+\delta'.
$$
Recall that numbers $C_{\tau}$ and $C_{\tau'}$ are not arbitrary.
They obey the relation (Legendre's identity \cite{Akh})
$$
C_{\tau}\tau'-C_{\tau'}\tau=1
$$
that originates from integrating ${\cal E}_{C_{\tau},C_{\tau'}}(z)$ around
the pole at zero point. Therefore, we get
\begin{equation}
\delta\tau'-\delta'\tau=0.
\label{arch}
\end{equation}
The only entire function with (\ref{as}) is $\psi=\alpha z+\beta$,
$\alpha, \beta\in {\bf C}$.
We can always choose $\alpha$ $(\alpha=-\delta/\tau)$ to put $\delta=0$,
then from (\ref{arch}) it follows that $\delta'=0$ also.
Hence, any two solutions of (\ref{sol2}) are related as
$$
{\cal E}(z)=\tilde{{\cal E}}(z)+\alpha z + \beta.
$$
The Weierstrass $\zeta$-function
\begin{equation}
\zeta(z)=\frac{1}{z}+\sum_{{\bf Z}}
\left(
\frac{1}{z-\omega_{nm}}+\frac{1}{\omega_{nm}}+\frac{z}{\omega_{nm}^2}
\right),~~\omega_{nm}=n\tau+m\tau',
\label{weier}
\end{equation}
satisfies the properties listed above and therefore gives a
peculiar solution of (\ref{sol2}). Thus, we have proved the following
\begin{lemma}
Any meromorphic function ${\cal E}(z)$ with only simple poles
at the points ${\bf Z}\tau+{\bf Z}\tau'$
with the residues $1/2\pi i$ and satisfying
(\ref{ee}) is of the form
\begin{equation}
{\cal E}(z)=\frac{1}{2\pi i}\zeta(z)+\alpha z + \beta
\label{gs}
\end{equation}
Clearly, when $\beta=0$ these functions are odd
${\cal E}(-z)=-{\cal E}(z)$.
\end{lemma}
Taking into account this proposition we can write
down a general solution of $\bar{\partial}x_i(z,\bar{z})=\frac{1}{k}
(t_i-v_i(z,\bar{z}))$
\begin{equation}
x_{diag}(z,\bar{z})=
\frac{1}{k}\int_{\Sigma_{\tau}}
d\bar{\eta} d\eta~~{\cal E}(z-\eta)
(t_i-v_i(\eta,\bar{\eta}))E_{ii}+h,
\label{gs1}
\end{equation}
where $h$ is a diagonal matrix in $sl(n,{\bf C})$.
Let us require this solution to be periodic, i.e.
$\psi_{\tau}=\psi_{\tau'}=0$. This determines the unknown diagonal part $t$:
\begin{equation}
t=\frac{1}{2i\Sigma_{\tau}}
\label{gs2}\int_{\Sigma_{\tau}}d\bar{\eta} d\eta~~v_i(\eta,\bar{\eta})E_{ii}.
\label{diag}
\end{equation}

Now we turn to the equation
\begin{equation}
\bar{\partial}x_{\alpha}(z,\bar{z})+\frac{\alpha(D)}{k}
x_{\alpha}(z,\bar{z})=-\frac{1}{k}v_{\alpha}(z,\bar{z}).
\label{dg}
\end{equation}
First we find the fundamental solution ${\cal E}_\alpha$
\begin{equation}
\bar{\partial}{\cal E}_{\alpha}(z)+\frac{\alpha(D)}{k}
{\cal E}_{\alpha}(z)=\delta(z,\bar{z}).
\label{dg0}
\end{equation}
Writing
${\cal E}_{\alpha}(z)=
e^{\frac{\alpha(D)}{k}(z-\bar{z})}Q_{\alpha}(z,\bar{z})$,
we get for $Q_{\alpha}(z,\bar{z})$ the following equation
$$
\bar{\partial}Q_{\alpha}(z,\bar{z})=e^{-\frac{\alpha(D)}{k}(z-\bar{z})}
\delta(z,\bar{z})=\delta(z,\bar{z})
$$
that tells us that $Q_{\alpha}(z,\bar{z})$ is a meromorphic function
having the simple pole at $z=0$ with the residue equal to $1/2\pi i$.
Assuming the fundamental solution ${\cal E}_{\alpha}(z)$ to be twice
periodic we find immediately the monodromy properties of
$Q_{\alpha}(z,\bar{z})$:
\begin{equation}
\begin{array}{l}
Q_{\alpha}(z+1)=Q_{\alpha}(z),\\
Q_{\alpha}(z+\tau)=e^{-\frac{\alpha(D)}{k}(\tau-\bar{\tau})}
Q_{\alpha}(z).
\end{array}
\end{equation}
The solution of the last problem is unique and is given by
$$
{\cal E}_{\alpha}(z)=
\frac{e^{\frac{\alpha(D)}{k}(z-\bar{z})}}{2\pi i}
\frac{\theta_{11}(z+\frac{\alpha(D)}{\pi k}\mbox{Im} \tau)\theta'_{11}(0)}
{\theta_{11}(z)\theta_{11}(\frac{\alpha(D)}{\pi k}\mbox{Im} \tau)},
$$
where
$$
\theta_{11}(z,\tau)=\sum_{n\in{\bf Z}}
e^{i\pi \tau\left(n+\frac{1}{2}\right)^2+2\pi i\left(n+\frac{1}{2}\right)
\left(z+\frac{1}{2}\right)}
$$
is the Jacobi $\theta$-function with monodromy properties
$$
\theta_{11}(z+1,\tau)=-\theta_{11}(z,\tau),~~
\theta_{11}(z+\tau,\tau)=-e^{-i\pi \tau -2\pi i z}\theta_{11}(z,\tau).
$$
Combining all pieces of our analysis together we can write
a general solution of the factorization problem for $sl(n,{\bf C})$ connection
\begin{equation}
X(z)=
\frac{1}{k}
\int_{\Sigma_{\tau}}d\bar{\eta} d\eta~~
{\cal E}(z-\eta)
(t_i-v_i(\eta,\bar{\eta}))E_{ii}+h
\label{gen1}
\end{equation}
\begin{equation}
-\frac{1}{k}\sum_{i\neq j}\int_{\Sigma_{\tau}}d\bar{\eta} d\eta~~
\frac{e^{\frac{\alpha_{ij}(D)}{k}((z-\eta)-(\bar{z}-\bar{\eta}))}}{2\pi i}
\frac{\theta_{11}(z-\eta+\frac{\alpha_{ij}(D)}{\pi k}\mbox{Im}
\tau)\theta'_{11}(0)}
{\theta_{11}(z-\eta)\theta_{11}(\frac{\alpha_{ij}(D)}{\pi k}\mbox{Im} \tau)}
v_{ij}(\eta,\bar{\eta})~E_{ij}.
\label{gen2}
\end{equation}
Now the problem is to fix the undetermined matrix $h$ by using
the boundary condition for  $X(0)$.

Recall \cite{GNH} that in the elliptic case one should choose the
following representative $J$ on the coadjoint $sl(n,{\bf C})$ orbit
\begin{equation}
J=1-u\otimes s^{\dagger},
\label{JJ}
\end{equation}
where $u,s$ are some vectors in ${\bf C}^n$. The requirement of vanishing
the diagonal entries of $J$ fixes the choice of components $s_i$:
$s_i^*=1/u_i$. Every solution $(\phi,A)$ of (\ref{const}) can
be brought to the form $(L,D)$, where $D$ is a constant diagonal matrix
and $L$ is the $L$-operator
\begin{equation}
L=\sum_{i}p_iE_{ii}-\frac{\nu}{2\pi
i} \sum_{i\neq j} e^{\frac{\alpha_{ij}(D)}{k}(z-\bar{z})}\frac{u_i}{u_j}
\frac{\theta_{11}(z+\frac{\alpha_{ij}(D)}{\pi k}\mbox{Im} \tau)\theta'_{11}(0)}
{\theta_{11}(z)\theta_{11}(\frac{\alpha_{ij}(D)}{\pi k}\mbox{Im} \tau)}E_{ij}
\label{lop}
\end{equation}
providing the Lax representation for the elliptic Calogero system \cite{GNH}.
Below we point out the connection of (\ref{lop}) with the $L$-operator
found by Krichever \cite{Kr}.

The Lie algebra ${\cal H}$ of the isotropy group of $J$ in $sl(n,{\bf C})$
is determined by the equation $[X,J]=0$, $X\in {\cal H}$ that is
equivalent to
\begin{equation}
u_i (s^{\dagger}X)_j-(Xu)_i s_j^{*}=0.
\label{dfg} \end{equation} Choosing in (\ref{dfg}) $i=j$ one gets
$(s^{\dagger}X)_i=\frac{s^{*}_i}{u_i}(Xu)_i$ and thereby (\ref{dfg})
reduces to $\frac{(Xu)_i}{u_i}=\frac{(Xu)_j}{u_j}=\lambda$, where
$\lambda\in {\bf C}$. One also has
$$(s^{\dagger}X)_i=\frac{s^{*}_i}{u_i}\lambda u_i=\lambda s^{*}_i= \lambda
(s^{\dagger})_i$$.  Thus, we find $\cal H$:
\begin{equation}
{\cal H}=\{X\in sl(n,{\bf C}):~~Xu=\lambda u,~~s^{\dagger}X=\lambda
s^{\dagger},~~\lambda\in {\bf C}\}.
\label{dfg1}
\end{equation}
{}From (\ref{dfg1}) we can read off that ${\cal H} \cap {\cal B}=0$, where
$\cal B$ is a maximal torus of $sl(n,{\bf C})$ (diagonal matrices).  Since
the real dimension of $sl(n,{\bf C})$ is $2(n^2-1)$ and ${\cal H}$ is
defined by $4n-4$ equations, we get $\dim {\cal
H}=2(n^2-1)-(4n-4)=2(n-1)^2$.

Now consider the decomposition of ${\cal G}=sl(n,{\bf C})$ in the direct
sum $$ {\cal G}={\cal H} \oplus {\cal B}\oplus {\cal C}, $$
where $\cal C$ is defined as an orthogonal subspace to ${\cal H} \oplus
{\cal B}$ with respect to the Killing metric $(X,Y)=\mbox{Tr}(XY)$. To
describe $\cal C$ explicitly we introduce a matrix $C$
$$ C=Z\otimes s^{\dagger}-u\otimes Y^{\dagger} $$
depending on two vectors $Z,Y\in {\bf
C}^n$. Let $X\in {\cal H}$ and $Xu=s^{\dagger}X=0$, then
$\mbox{Tr}(XC)=\mbox{Tr}(Z\otimes s^{\dagger}X-Xu\otimes Y^{\dagger})=0$.
On the other hand for $X$ arbitrary we have
$$
\mbox{Tr}(XC)=\sum_{ik}(z_i s_k^*-u_i y_k^*)x_{ki}
$$
and therefore for $C$ orthogonal to any
element $X=(x_i\delta_{ij})\in {\cal B}$ we have
$z_i s_i^*-u_i y_i^*=\beta$ for any $i$, where $\beta$ is an arbitrary
complex number. Orthogonality of ${\cal H}$  and $C$ also implies the
fulfillment of
$$
\mbox{Tr}(JC)=\beta n(1-<s^{\dagger}, u>)=\beta n(1-n)=0
$$
that gives $\beta=0$. Thus, $C$ is an element of ${\cal C}$ if
$y_{i}^*=\frac{s_i^*}{u_i}z_i=\frac{1}{u_i^2}z_i$. We also put
$\sum_{i} \frac{z_i}{u_i}=0$ to have the correct dimension of
${\cal C}$:  $\dim {\cal C}=2(n-1)$. This completes the description of
$\cal C$.

{\it Remark.} Just as for $su(n)$ case \cite{ABT} one can prove that $\cal
B$ and $\cal C$ form a pair of complementary Lagrangian subspaces with
respect to the symplectic form $\Xi(X,Y)=<J,[X,Y]>$ defined on ${\cal
B}\oplus {\cal C}$.
\vskip 0.5cm

Now one can easily find that $(Cu)_i=n z_i$ and
$(s^{\dagger}C)_i=-\frac{n}{u_i^2}z_i$.
This allows us to describe the action of a generic element
$X\in {\cal H} \oplus {\cal C}$ on the vectors $u$ and $s$:
\begin{equation}
\begin{array}{l}
(Xu)_i=\lambda u_i+n z_i,\\
(s^{\dagger}X)_i=\frac{\lambda}{ u_i}-\frac{n z_i}{u_i^2}.
\end{array}
\label{sd}
\end{equation}
Summing up the second lines in (\ref{sd}) and taking into account
$\sum \frac{z_i}{u_i}=0$, we find $\lambda$:
$$
\lambda=\frac{1}{n}\sum_i\frac{(Xu)_i}{u_i}.
$$
Solving (\ref{sd}) for $z_i$, we arrive at
\begin{lemma}
Let $X$ be an arbitrary element of ${\cal H}_J \oplus {\cal C}$.
Then the following relation
\begin{equation}
u_i(s^{\dagger}X)_i + (Xu)_i \frac{1}{u_i}
=\frac{2}{n}\sum_j\frac{(Xu)_j}{u_j}
\label{bas}
\end{equation}
is valid for any $i$.
\end{lemma}
Note that we can rewrite (\ref{bas}) as follows:
\begin{equation}
(s^{\dagger}X)_i u_i +
\sum_j\left(\delta_{ij}-\frac{2}{n}\right)\frac{(Xu)_j}{u_j}=0.
\label{bas1}
\end{equation}

We use Proposition 2 to fix the element $h$ in (\ref{gen2}). To this end
we put $X(0)\in {\cal H} \oplus {\cal C}$. Let us show that this
requirement determines $h$ completely. To simplify the calculations
we introduce
\begin{equation}
w_{ij}(\eta,\bar{\eta})=
\frac{e^{-\frac{\alpha_{ij}(D)}{k}(\eta-\bar{\eta})}}{2\pi i}
\frac{\theta_{11}(\eta-\frac{\alpha_{ij}(D)}{\pi k}\mbox{Im}
\tau)\theta'_{11}(0)}
{\theta_{11}(\eta)\theta_{11}(\frac{\alpha_{ij}(D)}{\pi k}\mbox{Im} \tau)},
\label{ww}
\end{equation}
and
\begin{equation}
q_{i}(\eta,\bar{\eta})=t_i-v_i(\eta,\bar{\eta}),
\label{qq}
\end{equation}
so that
\begin{equation}
X\equiv X(0)=
\frac{1}{k}\int_{\Sigma_{\tau}}d\bar{\eta} d\eta~~
{\cal E}(-\eta)q_i(\eta,\bar{\eta})E_{ii}+h
\label{null}
\end{equation}
$$
-\frac{1}{k}\sum_{i\neq j}\int_{\Sigma_{\tau}}d\bar{\eta} d\eta~~
w_{ij}(\eta,\bar{\eta})v_{ij}(\eta,\bar{\eta})E_{ij}.
$$
Using the explicit form of $X$ we calculate
\begin{eqnarray}
(Xu)_i & = &
\frac{1}{k}\int_{\Sigma_{\tau}}d\bar{\eta} d\eta~~
{\cal E}(-\eta)q_i u_i-
\frac{1}{k}\sum_{i\neq j}\int_{\Sigma_{\tau}}d\bar{\eta} d\eta~~
w_{ij}v_{ij}u_j+h_iu_i,   \label{om1}     \\
(s^{\dagger}X)_i & = &
\frac{1}{k}\int_{\Sigma_{\tau}}d\bar{\eta} d\eta~~
{\cal E}(-\eta)q_i \frac{1}{u_i}-
\frac{1}{k}\sum_{i\neq j}\int_{\Sigma_{\tau}}d\bar{\eta} d\eta~~
\frac{1}{u_j}w_{ji}v_{ji}+h_i\frac{1}{u_i}.
\label{om2} \\
\end{eqnarray}
With the help of (\ref{om1}) we also find
\begin{equation}
\sum_i \frac{(Xu)_i}{u_i} =
-\frac{1}{k}\sum_{i\neq j}\int_{\Sigma_{\tau}}d\bar{\eta} d\eta~~
\frac{u_j}{u_i}w_{ij}v_{ij}.
\label{om3}
\end{equation}
Substitution of (\ref{om1})-(\ref{om3}) in (\ref{bas}) results in
\begin{equation}
\frac{2}{k}\int_{\Sigma_{\tau}}d\bar{\eta} d\eta~~
{\cal E}(-\eta)q_i -
\frac{1}{k}\sum_{i\neq j}\int_{\Sigma_{\tau}}d\bar{\eta} d\eta~~
(\frac{u_j}{u_i}w_{ij}v_{ij}+\frac{u_i}{u_j}w_{ji}v_{ji})+ 2h_i=
\frac{2}{n}\sum_j \frac{(Xu)_j}{u_j}
\label{was}
\end{equation}
for any $i$. This equation allows us to find $h$:
\begin{equation}
h=-\frac{1}{k}\int_{\Sigma_{\tau}}d\bar{\eta} d\eta~~
{\cal E}(-\eta)\sum_{i}q_i E_{ii}+
\frac{1}{2k}\sum_{i\neq j}\int_{\Sigma_{\tau}}d\bar{\eta} d\eta~~
\left(\frac{u_j}{u_i}w_{ij}v_{ij}+\frac{u_i}{u_j}w_{ji}v_{ji}\right)
\left(E_{ii}-\frac{1}{n}I\right).
\label{fin}
\end{equation}
Thus, we arrive at
\begin{lemma}
A general solution $X(z,\bar{z})$ of the factorization problem for
$sl(n,{\bf C})$ connection satisfying $X(0)\in{\cal H} \oplus {\cal C}$
has the form
\begin{equation}
X(z,\bar{z}) =\frac{1}{k}\sum_{i}\int_{\Sigma_{\tau}}d\eta d\bar{\eta}~~
({\cal E}(z-\eta)-{\cal E}(-\eta))(t_i-v_i(\eta,\bar{\eta}))E_{ii}
\label{good}
\end{equation}
$$
-\frac{1}{k}\sum_{i\neq j}\int_{\Sigma_{\tau}}d\bar{\eta} d\eta~~
w_{ij}(\eta-z,\bar{\eta}-\bar{z})v_{ij}(\eta,\bar{\eta})E_{ij}
$$
$$
+\frac{1}{2k}\sum_{i\neq j}\int_{\Sigma_{\tau}}d\bar{\eta} d\eta~~
\left(\frac{u_j}{u_i}w_{ij}(\eta,\bar{\eta})v_{ij}(\eta,\bar{\eta})+
\frac{u_i}{u_j}w_{ji}(\eta,\bar{\eta})v_{ji}(\eta,\bar{\eta})\right)
\left(E_{ii}-\frac{1}{n}I\right),
$$
where $t$ and $w_{ij}$ are given by (\ref{diag}) and (\ref{ww}) respectively.
\end{lemma}
Using the explicit form (\ref{gs})
of ${\cal E}(z)$ and taking into account (\ref{diag}) it is easy to find
that the first line in (\ref{good}) reduces to
$$
\frac{\Phi(z,\bar{z})}{2ik \Sigma_{\tau}}
\sum_{i}\int_{\Sigma_{\tau}} d\bar{\eta}d\eta~~v_i(\eta,\bar{\eta})E_{ii}-
\frac{1}{2\pi ik}\sum_{i}\int_{\Sigma_{\tau}} d\bar{\eta}d\eta~~
(\zeta(z-\eta)+\zeta(\eta))v_i(\eta,\bar{\eta})E_{ii},
$$
where we have introduced a function
$$
\Phi(z,\bar{z})=
\int_{\Sigma_{\tau}}\frac{d\bar{\eta}d\eta}{2\pi i}~
(\zeta(z-\eta)+\zeta(\eta)).
$$
Hence, despite the function $\alpha z+\beta$ enters the fundamental
solution (\ref{gs}) solution (\ref{fac}) of the factorization problem does
not depend on it. Just as in the trigonometric case we get from
(\ref{good}) the following

\begin{theor}
The $R$-matrix
corresponding to $L$-operator (\ref{lop}) is the following matrix function
on $\Sigma_{\tau}\times \Sigma_{\tau}$
\begin{eqnarray}
R(z,\eta) & =  & \frac{\Phi(z,\bar{z})}{2ik\Sigma_\tau}E_{ii}\otimes E_{ii}
-\frac{1}{2\pi i k}(\zeta(z-\eta)+\zeta(\eta))\sum_i E_{ii}\otimes
E_{ii} \\   \label{gold}
& &
-\frac{1}{k}\sum_{i\neq j}
w_{ij}(\eta-z,\bar{\eta}-\bar{z})~E_{ij}\otimes E_{ji} \nonumber \\
& &
+\frac{1}{2k}\sum_{i\neq j}\left(E_{ii}-\frac{1}{n}I\right)\otimes
\left(\frac{u_j}{u_i}w_{ij}(\eta,\bar{\eta})~E_{ji}+
\frac{u_i}{u_j}w_{ji}(\eta,\bar{\eta})~E_{ij}\right). \nonumber \\
\end{eqnarray}
\end{theor}

{\it Remark}. $L$-operator as well as $R$-matrix (\ref{gold})
depends on vector $u\in {\bf C}^n$. However, by conjugating $L$
with a matrix $Q=e^U$, $U_{ij}=u_i\delta_{ij}$ this dependence may be
removed. The corresponding $R$-matrix is given by (\ref{gold})
with all $u_i=1$.
\vskip 0.5cm

Now we are going to make a connection with the Sklyanin result \cite{Skl}.
Without loss of generality we can assume that the integration domain
$\Sigma_{\tau}$ has vertexes at points $\pm\frac{1}{2}\pm\frac{\tau}{2}$.
Then by the oddness of $\zeta$-function one has
$\int_{\Sigma_{\tau}}d\eta d\bar{\eta}~\zeta(\eta)=0$ and
therefore $\Phi(z,\bar{z})$ reduces to
\begin{equation}
\Phi(z,\bar{z})=\int_{\Sigma_{\tau}}\frac{d\bar{\eta}d\eta}{2\pi
i}~\zeta(z-\eta).
\label{odd}
\end{equation}
Eq.(\ref{odd}) means that $\Phi(z,\bar{z})$ is a solution of the equation
$\bar{\partial}\Phi(z,\bar{z})=1$, i.e. $\Phi(z,\bar{z})=\bar{z}+f(z)$,
where $f(z)$ is an entire function. The monodromy properties of $\zeta$
define the ones for $\Phi(z,\bar{z})$:
$$
\Phi(z+\tau,,\bar{z}+\bar{\tau})=\Phi(z,\bar{z})+ \frac{\Sigma _{\tau}}{\pi}
C_{\tau},~~
$$
where $\tau$ denotes any of two periods $1$ and $\tau$. For
$f(z)$ the equation above implies that
\begin{equation}
f(z+\tau)-f(z)=\frac{\Sigma _{\tau}}{\pi} C_{\tau}-\bar{\tau}.
\label{fun}
\end{equation}
The only entire function obeying (\ref{fun}) is $f(z)=\alpha z+\beta$ with
$\alpha=\frac{C_{\tau}/\pi-\bar{\tau}}{\tau}$. However, we have to prove
that $\alpha$ is defined by the {\it same} formula for both periods
$1$ and $\tau$. In other words, we have to prove the relation
\begin{equation}
\frac{C_{\tau}/\pi-\bar{\tau}}{\tau}=\frac{C_1/\pi-1}{1},
\label{rel}
\end{equation}
where $C_1$ corresponds to the shift of $\zeta$-function along the period
1. Fortunately, (\ref{rel}) is satisfied since it reduces to the identity
$\mbox{Im}\tau=\frac{\tau-\bar{\tau}}{2i}=
\frac{C_1\tau-C_{\tau}1}{2\pi i}\Sigma_{\tau}$ that follows in its turn
from Legendre's identity on the numbers $C_{\tau}$, $C_{\tau'}$.
The constant $\beta$ is equal to zero by the oddness
of $\Phi(z)$. Thus, we get for $\Phi(z,\bar{z})$
the following explicit answer
\begin{equation}
\Phi(z,\bar{z})=\bar{z}-z+\frac{C_1}{\pi}\mbox{Im}\tau~z.
\label{ans}
\end{equation}
In \cite{Skl} Krichever's $L$-operator \cite{Kr}:
\begin{equation}
L^{Kr}=\sum_{i}p_iE_{ii}-\frac{\nu}{2\pi i}
\sum_{i\neq j}
\frac{\sigma(z+\frac{\alpha_{ij}(D)}{\pi k}{\mbox Im \tau})}
{\sigma(z)\sigma(\frac{\alpha_{ij}(D)}{\pi k}{\mbox Im \tau})}E_{ij}
\label{lkr}
\end{equation}
was used to find the corresponding $R$-matrix. Due to the identity
$$
G_{ij}(z)=\frac{\sigma(z+\frac{\alpha_{ij}(D)}{\pi k}\mbox{Im}\tau)}
{\sigma(z)\sigma(\frac{\alpha_{ij}(D)}{\pi k}\mbox{Im}\tau)}=
\exp{\left(\frac{C_1\alpha_{ij}(D)\mbox{Im}\tau}{\pi k} z\right)}
\frac{\theta_{11}(z+\frac{\alpha_{ij}(D)}{\pi k}\mbox{Im} \tau)\theta'_{11}(0)}
{\theta_{11}(z)\theta_{11}(\frac{\alpha_{ij}(D)}{\pi k}\mbox{Im} \tau)}
$$
it is easy to see that $L^{Kr}$ is related to (\ref{lop}) by the similarity
transformation
\begin{equation}
L^{Kr}(z)=Q(z,\bar{z})L(z,\bar{z})Q(z,\bar{z})^{-1},
\label{sim}
\end{equation}
where $Q(z,\bar{z})=e^{\frac{D}{k}\Phi(z,\bar{z})}$. Note also, that
$L^{Kr}(z)$ does not depend on $\bar{z}$. The Poisson bracket for $L^{Kr}$
is obtained in the usual way
$$
\{L^{Kr}_1,L^{Kr}_2\}=[Q_1\{L_1,Q_2\}Q_1^{-1}Q_2^{-1},L^{Kr}_2]+
[Q_2\{Q_1,L_2\}Q_1^{-1}Q_2^{-1},L^{Kr}_1]+
$$
$$
Q_1Q_2\{L_1,L_2\}Q_1^{-1}Q_2^{-1},
$$
where the omitted spectral parameters $z$ and $\eta$ can be easily
restored. Calculating $\{L_1,Q_2\}$ with the help of the canonically
conjugated variables
$\{P,D\}=\frac{1}{2i\Sigma_{\tau}}\sum_iE_{ii}\otimes E_{ii}$,
$P=\sum_i p_iE_{ii}$ on the reduced phase space, we recover the
$R$-matrix for $L^{Kr}$:
\begin{equation}
R(z,\eta) =
-\frac{1}{2\pi i k}(\zeta(z-\eta)+\zeta(\eta))\sum_i E_{ii}\otimes E_{ii}
\label{gold1}
\end{equation}
$$
-\frac{1}{2\pi ik}\sum_{i\neq j}
G_{ij}(z-\eta)~E_{ij}\otimes E_{ji}
-\frac{1}{4\pi ik}\sum_{i\neq j}G_{ij}(\eta)
\left(E_{ii}+E_{jj}-\frac{1}{n}I\right)\otimes E_{ij}
$$
that is precisely the result of \cite{Skl}.
$$~$$
{\bf ACKNOWLEDGMENT}
One of the authors (G.A.) is grateful to L.Chekchov, A.Marshakov,
S.Frolov for useful discussions and especially to A.Gorsky for
illuminating the unsolved problems in the theory of integrable many body
systems.  This work is supported in part by RFFR under grant N93-011-147
and by ISF under grants M1L-000 and M1L-300.
$$~$$
{\large \bf APPENDIX}
\appendix
\section{Proof of Proposition 5.}
\setcounter{equation}{0}

Recalling the explicit form of the $L$-operator we first compute
$$
\sum_{i\neq j}\frac{e^{-\frac{i\alpha_{ij}(X)}{k}(\varphi-\varphi')}
e^{\frac{i\alpha_{kl}(X)}{k}(\pi-\varphi)}}
{\sin{\frac{\pi\alpha_{kl}(X)}{k}}}[E_{ij},E_{kl}] \otimes E_{ji}=
$$
$$
\sum_{\stackrel{i\neq j}{l\neq j}}
\frac{e^{-\frac{i\alpha_{ij}(X)}{k}(\varphi-\varphi')}
e^{\frac{i\alpha_{jl}(X)}{k}(\pi-\varphi)}}
{\sin{\frac{\pi\alpha_{jl}(X)}{k}}}E_{il}\otimes E_{ji}-
\sum_{\stackrel{i\neq j}{k\neq i}}
\frac{e^{-\frac{i\alpha_{ij}(X)}{k}(\varphi-\varphi')}
e^{\frac{i\alpha_{ki}(X)}{k}(\pi-\varphi)}}
{\sin{\frac{\pi\alpha_{ki}(X)}{k}}}E_{kj}\otimes E_{ji}.
$$
The preceding sums can be divided in four parts
\begin{equation}
\sum_{i\neq j}
\frac{e^{-\frac{i\alpha_{ij}(X)}{k}(\varphi-\varphi')}
e^{\frac{i\alpha_{ji}(X)}{k}(\pi-\varphi)}}
{\sin{\frac{\pi\alpha_{ji}(X)}{k}}}E_{ii}\otimes E_{ji}-
\label{dg1}
\end{equation}
\begin{equation}
\sum_{i\neq j}
\frac{e^{-\frac{i\alpha_{ij}(X)}{k}(\varphi-\varphi')}
e^{\frac{i\alpha_{ji}(X)}{k}(\pi-\varphi)}}
{\sin{\frac{\pi\alpha_{ji}(X)}{k}}}E_{jj}\otimes E_{ji}+
\label{dg2}
\end{equation}
\begin{equation}
\sum_{i\neq j\neq l}
\frac{e^{-\frac{i\alpha_{ij}(X)}{k}(\varphi-\varphi')}
e^{\frac{i\alpha_{jl}(X)}{k}(\pi-\varphi)}}
{\sin{\frac{\pi\alpha_{jl}(X)}{k}}}E_{il}\otimes E_{ji}-
\label{dg3}
\end{equation}
\begin{equation}
\sum_{i\neq j\neq k}
\frac{e^{
-\frac{i\alpha_{ij}(X)}{k}(\varphi-\varphi')}
e^{\frac{i\alpha_{ki}(X)}{k}(\pi-\varphi)}}
{\sin{\frac{\pi\alpha_{ki}(X)}{k}}}E_{kj}\otimes E_{ji}.
\label{dg4}
\end{equation}

Taking into account the relations
$$
-\alpha_{ij}(X)(\varphi-\varphi')+\alpha_{jl}(X)(\pi-\varphi)=
\pi \alpha_{jl}(X)+\alpha_{ij}(X)\varphi'+\alpha_{li}(X)\varphi,
$$
$$
-\alpha_{ij}(X)(\varphi-\varphi')+\alpha_{ki}(X)(\pi-\varphi)=
\pi \alpha_{ki}(X)+
\alpha_{ij}(X)\varphi'+\alpha_{jk}(X)\varphi,
$$
and combining (\ref{dg1}) and (\ref{dg2}), we finally get
\begin{eqnarray}
-\sum_{i\neq j}
\frac{e^{-\frac{i\alpha_{ij}(X)}{k}(\pi-\varphi')}}
{\sin{\frac{\pi\alpha_{ji}(X)}{k}}}
\left(E_{ii}-E_{jj}\right)\otimes E_{ji}~~~  \label{rom} \\
+\sum_{i\neq j\neq l}
\frac{e^{i
\frac{\pi \alpha_{jl}(X)+\alpha_{ij}(X)\varphi'+\alpha_{li}(X)\varphi}{k}}}
{\sin{\frac{\pi\alpha_{jl}(X)}{k}}}
E_{il}\otimes E_{ji}~~~\mbox{(A)}   \label{A} \\
-\sum_{i\neq j\neq l}
\frac{e^{i
\frac{\pi \alpha_{li}(X)+\alpha_{ij}(X)\varphi'+\alpha_{jl}(X)\varphi}{k}}}
{\sin{\frac{\pi\alpha_{li}(X)}{k}}}
E_{lj}\otimes E_{ji}.~~~\mbox{(B)}  \label{B} \\
\end{eqnarray}
Now we calculate the commutator
$$
\left[
\sum_{k}E_{kk}\otimes E_{kk},
I\otimes \sum_{i\neq j}\frac{e^{i\frac{\alpha_{ij}(X)}{k}(\pi-\varphi')}}
{\sin{\frac{\pi\alpha_{ij}(X)}{k}}}E_{ij}
\right]=
$$
$$
\sum_{i\neq j}
\frac{e^{\frac{i\alpha_{ij}(X)}{k}(\pi-\varphi')}}
{\sin{\frac{\pi\alpha_{ji}(X)}{k}}}
\left(E_{ii}-E_{jj}\right)\otimes E_{ij}
$$
Combining the previous expression with (\ref{rom}), we obtain
$$
-\left(\sum_{i\neq j}
\frac{e^{\frac{-i\alpha_{ij}(X)}{k}(\pi-\varphi')}}
{\sin{\frac{\pi\alpha_{ji}(X)}{k}}}
\left(E_{ii}-E_{jj}\right)\otimes E_{ji}
\right)\left(\epsilon(\varphi-\varphi')+
\epsilon(\varphi'-\varphi)\right)=0.
$$
As to (\ref{A}) and (\ref{B}), they have the counterparts coming from the
second commutator in (\ref{qaq}). These counterparts can be immediately
derived by twisting the factors of the tensor product in
(\ref{A}) and (\ref{B}) and changing $\varphi$ for $\varphi'$. We
denote them by $C$ and $D$:
\begin{eqnarray}
-\sum_{i\neq j\neq l}
\frac{e^{i
\frac{\pi \alpha_{jl}(X)+\alpha_{ij}(X)\varphi+\alpha_{li}(X)\varphi'}{k}}}
{\sin{\frac{\pi\alpha_{jl}(X)}{k}}}
E_{ji}\otimes E_{il}.~~~\mbox{(C)}  \label{C} \\
-\sum_{i\neq j\neq l}
\frac{e^{i
\frac{\pi \alpha_{li}(X)+\alpha_{ij}(X)\varphi+\alpha_{jl}(X)\varphi'}{k}}}
{\sin{\frac{\pi\alpha_{li}(X)}{k}}}
E_{ji}\otimes E_{lj}.~~~\mbox{(D)}  \label{D}
\end{eqnarray}
The change of indices in (\ref{C}) and (\ref{D}) leads to $C=-B$ and $D=-A$.
Now subtracting $(C+D)\epsilon(\varphi'-\varphi)$ from
$(A+B)\epsilon(\varphi-\varphi')$, we get zero.
Continuing this line of reasoning, it is not hard to prove
that the moment part of the $L$-operator also
satisfies (\ref{qaq}).

\end{document}